# A new framework for X-ray absorption spectroscopy data analysis based on machine learning: XASDAML

**Xue Han[1,2#], Haodong Yao[1,3#], Fei Zhan[1], Xueqi Song[1], Junfang Zhao[2*], Haifeng Zhao[1*]**

[1] Multi-disciplinary Research Division, Institute of High Energy Physics, Chinese Academy of Sciences, Beijing 100049, China

[2] School of Science, China University of Geosciences, Beijing, 100083, China

[3] University of Chinese Academy of Sciences, Beijing 100049, China

# These authors contributed equally to the work.

* Corresponding author: zhaohf@ihep.ac.cn, jfzhao@cugb.edu.cn

## Highlights

1. An open-access framework that integrates the entire XAS data processing workflow based on machine learning—including dataset construction, data filtering, machine learning modeling, prediction, and model evaluation—into a unified and flexible platform, including data statistical analysis for comprehensive exploration of data features and trends.
2. Employs principal component analysis, clustering, and supervised machine learning models to explore structure–spectra relationships in large-scale XAS datasets. It also supports statistical analysis and visualization, aiding in the identification of potential patterns and relationships in the data.
3. Each module operates independently and is accessible via Jupyter Notebook, enabling users to easily replace or upgrade components to meet specific research needs and adapt to future advancements in XAS data analysis.
4. The application to copper datasets demonstrates its ability to efficiently handle large and complex data, providing accurate predictions and comprehensive model evaluations.

# Abstract

X-ray absorption spectroscopy (XAS) is a powerful technique to probe the electronic and structural properties of materials. With the rapid growth in both the volume and complexity of XAS datasets driven by advancements in synchrotron radiation facilities, there is an increasing demand for advanced computational tools capable of efficiently analyzing large-scale data. To address these needs, we introduce XASDAML,a flexible, machine learning based framework that integrates the entire data-processing workflow—including dataset construction for spectra and structural descriptors, data filtering, ML modeling, prediction, and model evaluation—into a unified platform. Additionally, it supports comprehensive statistical analysis, leveraging methods such as principal component analysis and clustering to reveal potential patterns and relationships within large datasets. Each module operates independently, allowing users to modify or upgrade modules in response to evolving research needs or technological advances. Moreover, the platform provides a user-friendly interface via Jupyter Notebook, making it accessible to researchers at varying levels of expertise. The versatility and effectiveness of XASDAML are exemplified by its application to a copper dataset, where it efficiently manages large and complex data, supports both supervised and unsupervised machine learning models, provides comprehensive statistics for structural descriptors, generates spectral plots, and accurately predicts coordination numbers and bond lengths. Furthermore, the platform streamlining the integration of XAS with machine learning and lowering the barriers to entry for new users.

# Introduction

X-ray absorption spectroscopy (XAS) has long been a fundamental technique available at synchrotron radiation facility, enabling the investigation of the electronic and structural properties of materials at the atomic scale. It provides critical insights into local geometric structure, oxidation states, coordination symmetry, and bond distances associated to the absorber[1].These atomic-level details are essential for understanding a wild range of material behaviors, including catalytic activity[2], structural properties[3], and phase transitions[4]. To facilitate the analysis of XAS data, various theoretical frameworks and calculation packages have been developed. Notable examples include ORCA (a quantum chemistry package based on density functional theory, including time-dependent extensions)[5], FEFF(a code based on the real-space multiple-scattering theory for XAS calculations)[6], MXAN (which uses full multiple-scattering theory to analyze Extended X-ray Absorption Fine Structure (EXAFS) and X-ray absorption near-edge structure (XANES) data)[7], and FDMNES (for finite difference method near rdge structure, uses the density functional theory)[8]help to simulate or fit XAS data from three-dimensional structural models. Some alternative schemes are also avaible for the data processing of time-resolved X-ray absorption spectroscopy based on the variant of previous packages[9].

With the development and operation of fourth generation synchrotron radiation facilities based on diffraction-limited storage rings, application of space resolved XAS measurement with nano-probe technology and time-dependent XAS with quick scans promote the volume of XAS datasets to a level that traditional data analysis methods struggle to process efficiently. Furthermore, the demand for real-time data processing at beamlines to extract relevant structural information and guide subsequent experimental measurements has created a need for tools that can handle large datasets quickly and accurately. The integration of machine learning (ML) and artificial intelligence(AI) into XAS analysis has emerged as a promising solution, particularly in automating batch data processing, improving feature extraction, and lowering the entry barrier for new users. This shift was catalyzed by Timoshenko et al., who first introduced supervised machine learning to refine the 3D geometry of metal catalysts from XANES spectroscopy, enabling the tracking of heterogeneous catalyst structures under operando conditions[10]. Since then, ML methods have proven valuable in enhancing the accuracy and scalability of XAS data interpretation. For instance, Timoshenko et al.[11]employed Principal Component Analysis (PCA) and neural networks to interpret EXAFS data, revealing how these tools can be used to enhance the interpretation of structural information and uncover new insights into material behavior. Machine learning has also been instrumental in linking spectral features to fundamental structural parameters. A.A. Guda et al. (2021) [12][13] demonstrated how ML techniques can correlate detailed spectral descriptors, such as edge position and intensity variations, with underlying structural features like coordination numbers, bond lengths, and oxidation states. By mapping these spectral features to atomic-scale metrics, their work highlighted ML's abilit to bridge the gap between raw XAS data and meaningful structural insights, which are critical for understanding and predicting material properties. These efforts suggest that ML models, including Multi-Layer Perceptron (MLP), support vector machines(SVM), convolutional neural networks(CNN), and random forests(RF), can help reveal complex structure-property relationships in challenging XAS datasets[12][14][15][16][17]. Such techniques are not only advancing the understanding of materials at the atomic level but are also addressing growing demands for high-throughput analysis and the need for automated interpretation of large-scale XAS data.

With the growing interest in integrating machine learning into XAS data analysis, several software tools combining XAS and ML have been developed. Among them, TRixs[18] by Torrisi et al., XANESNET [19] [20] by Madkhali et al., and PyFitIt [13] [21][22] developed by Guda's group each represent significant steps toward more automated workflows. While these platforms introduce important functionalities—such as data preprocessing, polynomial fitting, deep neural network training, and regression-based structural prediction—they also highlight areas where further refinement could be beneficial. For example, TRixs leverages random forests and a multiscale featurization approach (i.e., polynomial-fit features) to interpret XAS spectra of transition metal oxides, focusing on properties such as Bader charge and mean nearest neighbor distance. Although these techniques enhance model interpretability and automate data-driven analyses, the emphasis on polynomial fitting and a narrower material scope may limit its applicability in broader contexts. XANESNET, while offering detailed workflows for training and prediction, requires users to manually prepare and organize datasets—often sourced from external databases like the Materials Project—which can add extra steps for ensuring compatibility. Lastly, PyFitIt provides multiple regression algorithms for structural prediction, but its tightly integrated design and limited customization options may pose challenges, particularly when adapting the platform to newer computing environments or specific research needs.

In this study, we introduce XAS Data Analysis based on Machine Learning(XASDAML), an open-source framework designed to streamline the entire XAS data processing and analysis workflow. Building on existing solutions, XASDAML integrates all essential modules—from simulating XAS spectra using atomic coordinates of deformed materials and constructing machine learning models to predicting structural descriptors and evaluating model performance. This unified approach lowers the barrier for XAS researchers who may not have extensive ML expertise; with XASDAML, users can generate and curate their own XAS datasets that closely match their system of interest, specify the structure geometry they wish to study, select or develop an appropriate ML model, and obtain reliable structural predictions from newly measured XAS data. Beyond these core capabilities, XASDAML offers robust data visualization and statistical analysis toolkits, enabling users to gain deeper insights into spectral features and structural trends. Accessible via a user-friendly Jupyter Notebook interface, XASDAML consolidates spectral and structural calculations, data visualization, machine learning modeling, and predictive analysis into a single platform. This design not only enables users to customize parameters within each module but also lets them monitor program execution and review outputs in real time. Ultimately, XASDAML facilitates high-throughput, automated analysis and fosters broader adoption of ML-driven XAS research.

## Program description

XASDAML is a machine learning–based framework for XAS data analysis, consisting of 12 modules developed in Python. It is accessible through a Jupyter Notebook interface, offering an interactive environment that integrates code execution, data visualization, and documentation for efficient, reproducible research. XASDAML spans the entire XAS data analysis workflow, starting with the construction of training datasets through the simulation of XAS spectra and structural descriptors from structures. These simulated data are combined into feature/label datasets, which are then divided into training, validation, and test subsets after outlier filtering— facilitated by data plotting and statistical analysis. The framework then builds and trains machine learning models using the prepared datasets, applies these models to predict structural descriptors, and evaluates predictive performance (see Figure 1).

Most modules take the outputs of previous modules as inputs and generate files required by subsequent modules, as illustrated in Table 1. For clarity, we group the modules into four functional blocks: Dataset Calculation, Dataset Optimization, Machine Learning Modeling, and Prediction & Prediction Analysis. Additional modules, such as Data Visualization and Statistical Analysis, function as plug-ins that complement and enhance the core workflow. All modules are designed to operate independently, each with distinct inputs and outputs. They share a consistent architecture comprising five sections—introduction, library imports, parameter settings, function definitions, and the main program. The Parameter Settings section, accessible through the Jupyter Notebook interface (as shown in Figure 2), allows users to configure all necessary parameters, providing flexibility for customization and ensuring ease of use.

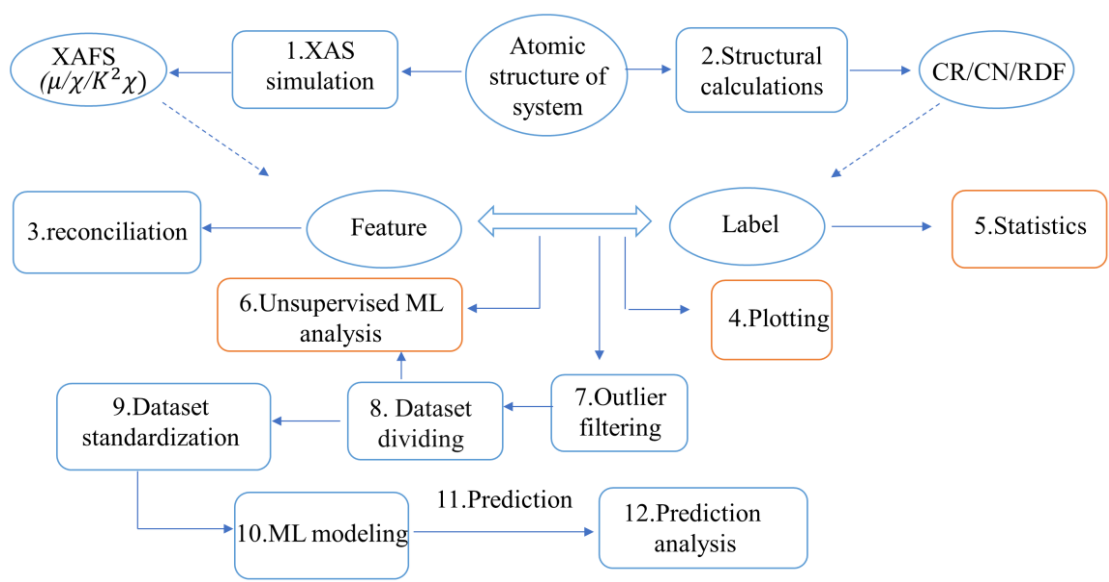

Figure 1. Flowchart illustrating the XASDAML-based workflow for XAS data analysis, from XAS calculations to model training and prediction analysis. The numbers on each module indicate the recommended execution order. See Table 1 for additional details on each module. CN refers to the coordination number of the first shell around the absorber, and CR denotes the average bond length in the absorber's first shell; RDF refers to the radial distribution function.

Table 1. 12 modules in XASDAML, including their inputs and outputs. The modules are grouped into four blocks—Dataset Calculation (Block 1), Dataset Optimization (Block 2), Machine Learning Modeling (Block 3), and Prediction & Prediction Analysis (Block 4)—as well as three supplementary toolkits. CN: the coordination number of the first shell, and CR: the average bond length in the absorber's first shell; RDF： the radial distribution function.

| No. of module | Functions | No. of block | Input | Output |
|---|---|---|---|---|
| 1 | Simulation of XAS | Block 1 | 3D atomic structure of material | Spectra($\mu,\chi$,wt) |
| 2 | Simulation of structure descriptors | Block1 | 3D atomic structure of material | Structure descriptors(CR, CN, RDF） |
| 3 | Data reconciliation | Block2 | Spectra($\mu$) | Spectra after interpolation |
| 4 | Spectra & structure descriptors plot | Toolkit1 | Spectra($\mu, \chi$, wt) & structure descriptors | Figures for spectra($\mu, \chi$, wt) & structure descriptors (RDF) |
| 5 | Statistics of structure descriptor | Toolkit2 | Structure descriptors | Tables for statistical analysis &figures for structure descriptors |
| 6 | Data analysis of spectra and structure descriptors | Toolkit3 | Spectra($\mu, \chi$, wt) & structure descriptors | Data analysis figures for spectra with structure descriptors |
| 7 | Dataset optimization | Block2 | Spectra($\mu, \chi$, wt) & structure descriptors | Spectra, structure descriptors and index of outliers' samples |
| 8 | Dataset division | Block2 | Spectra($\mu, \chi$, wt) & structure descriptors & index of outliers samples | Divided dataset (training set, validation set and test set) |
| 9 | Dataset standardization | Block2 | The divided dataset (training set, validation set and test set) | Dataset after normalization or PCA transformation |
| 10 | Machine learning modeling | Block3 | Dataset after normalization or PCA transformation | Optimal model & figure of loss curve |
| 11 | Prediction | Block4 | Optimal model & test set | Prediction of test set |
| 12 | Model performance evaluation | Block4 | Prediction, true labels, and features of test set | Statistics table of prediction value and true value& data analysis for prediction results |

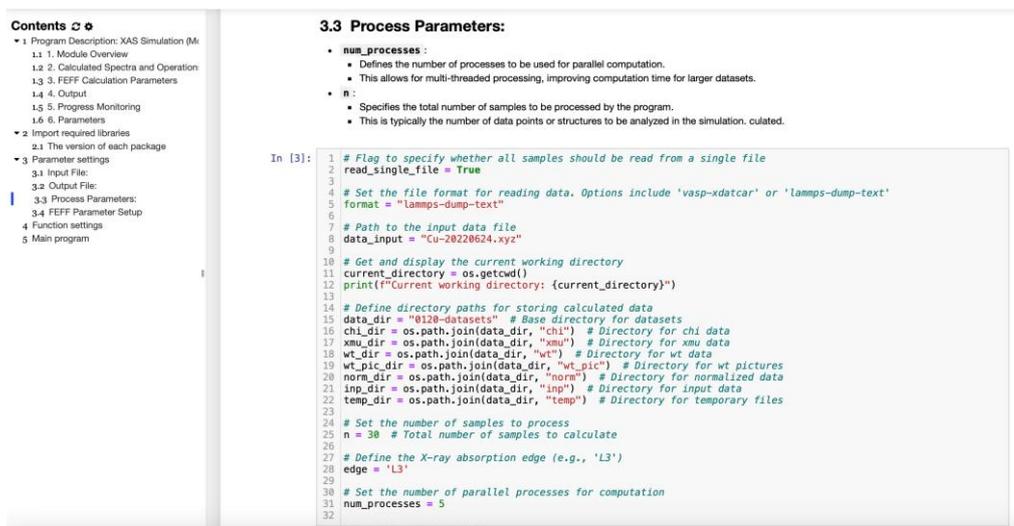

Figure 2. Screenshot of module 1(Simulation of XAS) open in jupyter notebook. On the left is the table of contents outlining the module's architecture, while the displays the code section. Here, the parameter settings section is shown, where users can configure necessary parameters and their descriptions to enable flexible analysis.

## Block 1: Dataset calculation (Module1 & 2)

XASDAML provides two modules for generating the features and labels required by machine learning models. Module 1 simulates X-ray absorption spectra, while Module 2 extracts structural descriptors, both working from the same input atomic coordinates. Although their outputs are saved by default in the same directory for convenience, each module operates independently, giving users the flexibility to run them separately or together depending on their needs.

Module 1 integrates the widely used FEFF package[23] for XAS simulations. After reading the input structure of clusters (potentially with different deformations), the module automatically generates standard FEFF input files. Users can customize simulation parameters as desired, and the module, by default, calculates the standard absorption coefficient $\mu(E)$, the normalized EXAFS function $\chi(k)$, where the abscissa at $\chi(E) = [\mu(E) - \mu_0(E)]/\mu_0(E)$ is converted from energy to wavenumber $k$, and the wavelet transform data $wt$. All these outputs serve as potential features for subsequent machine learning tasks. A progress bar and process window assist users in monitoring real-time simulation status—a valuable feature for large-scale datasets. Moreover, the module offers parallel processing options that can be tailored to the user's computational resources. Compared with other simulation tools [24],it is suitable for high-throughput calculation requirements.

Module 2 computes structural descriptors that are commonly analyzed in XAS studies, including CN, CR, and RDF. CN and CR often play pivotal roles in EXAFS fitting, whereas RDF provides a more comprehensive view of how atoms are distributed within a material. Additional descriptors such as geometric symmetry can be integrated with minimal effort, as each descriptor is calculated independently. To identify neighboring atoms, the module includes six algorithms— Effective Coordination Number Algorithm (EconNN), O'Keeffe's Voronoi Coordination Method (VoronoiNN), Crystal Nearest Neighbor Method (CrystalNN), Minimum Distance Algorithm (MinimumDistanceNN), Jmol Algorithm (JmolNN), and Brunner Method (BrunnerNN) [25] — allowing users to select the method best suited to their material system. As with Module 1, the computations are indexed to match the original sample order, a progress bar facilitates monitoring, and parallel processing is available to accelerate the workflow.

## Block 2: Dataset Optimization (Module 3, 7, 8, and 9)

Block 2 refines the spectra and structural descriptors produced in Block 1, ensuring that the data are of sufficient quality and consistency for training machine learning models in Block 3. This block reconciles simulated data (Module 3), filters outliers (Module 7), divides the dataset into training, validation, and test subsets (Module 8), and standardizes the data (Module 9). Together, these steps promote robust and flexible dataset management while enhancing overall model performance.

Module 3 interpolates spectra containing inconsistent energy points to preserve data smoothness and integrity. It also offers plotting tools that enable users to visually inspect their results and manually exclude any anomalous data. For a more direct overview of potential outliers, Module 4—discussed later—provides statistical summaries of all descriptors, making it easier to identify and isolate problematic samples.

Module 7 applies user-defined filters, such as thresholds on CN or CR, to eliminate unphysical data. For instance, if certain algorithms (e.g., JmolNN) occasionally assign a CN of 15 to copper atoms—an improbable value for well-organized Cu clusters—those samples can be flagged and removed. The program preserves the original dataset and logs the indices of filtered samples in a separate file, copying the excluded spectra and structural descriptors to a distinct directory. This approach not only streamlines iterative data exploration—allowing quick adjustments to filter parameters and reanalysis—but also retains full traceability of the original data.

Following data cleanup, Module 8 partitions the dataset into training, validation, and test sets according to user specifications (defaulting to a 7:2:1 ratio). To ensure efficient I/O, each set compiles both the spectra and structural descriptors into a single file. Module 9 then standardizes these subsets, handling data normalization and optional PCA transformations to remove discrepancies in feature scales. Key parameters (e.g., mean, standard deviation) are stored to maintain consistency in future predictions, making it straightforward to apply the same transformations to new or updated data.

## Block 3: Machine learning model(Module 10s)

Block 3 of XASDAML focuses on the supervised machine learning modeling process, providing a framework for training models, optimizing hyperparameters, and outputting the best-performing model for prediction. This is typically the most time-consuming stage in the XASDAML pipeline, second only to the spectral simulations performed in Module 1. The computational cost is largely determined by the chosen machine learning model, the model architecture, and the criteria for convergence.

XASDAML offers three widely-used machine learning model types for XAS data analysis: MLP, CNN, and RF. Both MLP and CNN support regression and classification tasks, while RF is limited to classification tasks. These models are implemented across several modules: Module 10-

1-1 (MLP-R) for regression with MLP, Module 10-1-2 (MLP-C) for classification with MLP, Module 10-2-1 (CNN-R) for regression with CNN, Module 10-2-2 (CNN-C) for classification with CNN, and Module 10-3 (RF-R) for regression with RF. Each model provides customizable hyperparameters, including the number of layers and nodes for MLP, the convolution kernel size and the number of channels for CNN, and the number of trees and tree depth for RF. These settings are designed to be user-friendly, allowing researchers to tailor the architecture to suit their specific dataset and research objectives. Additionally, training parameters, such as the number of epochs and learning rate, are adjustable to optimize model performance. To prevent overfitting and improve generalization, both MLP and CNN models support L2 regularization. The early stopping mechanism is integrated into the training process to save computational resources. This feature halts training when model performance no longer improves over several epochs, as monitored by the validation set's loss value. Additionally, model checkpointing ensures that the model parameters are saved at the optimal performance point, so the final model produced is the one with the best predictive accuracy.

The module also provides built-in tools to track and visualize the training dynamics, including loss curves and key performance metrics like mean squared error (MSE) and mean absolute error (MAE). This enables users to quickly identify issues such as overfitting or underfitting and adjust the model parameters accordingly. The combination of real-time feedback and intuitive visualization is aimed at helping novice users navigate the model training process effectively. For beginners, the module is designed to streamline the workflow, while offering sufficient flexibility for advanced users to customize their models for specific XAS data analysis needs.

## Block 4: Prediction & prediction analysis (Module 11& 12)

Block 4 focuses on the prediction phase, where the trained ML models are applied to predict structure descriptors from measured XAS spectra and subsequently evaluated for their performance. This block consists of two modules: Module 11, the prediction module, and Module 12, the prediction analysis module.

Module 11 importing a well-trained ML model and using it to predict structure descriptors based on input spectra. The module supports both single and batch input formats. It processes the spectra and outputs the predicted structure descriptors, but it does not assess the quality of these predictions. This separation allows users to focus on extracting the desired predictions before moving on to the evaluation stage.

Module 12, on the other hand, takes the output from Module 11 and evaluates the performance of the trained ML model. It compares the predicted values to the corresponding ground truth values (if available) and computes a range of common statistical indicators to assess model accuracy. Key evaluation metrics such as MSE, MAE, determination coefficient ($R^2$), accuracy, recall, F1 score, precision and Area Under the Curve (AUC)[26][27] are available for user to choose according to the pattern of model (regression or classification) and the data pattern. The module produces a comprehensive analysis of the prediction performance, including statistical tables of the prediction data and ground truth, as well as visual comparisons through parity plots, line graphs, histograms, and box plots of residual distributions. These visualization tools are invaluable for quickly identifying performance trends and potential discrepancies between predicted and observed values.

Module 12 also facilitates direct comparison of multiple machine learning models, making it useful for evaluating and selecting the best-performing model on the same dataset.

Additionally, Module 12 includes an optional feature for unsupervised machine learning analysis, providing K-means clustering and PCA to further explore the relationship between the input spectra and the model's prediction accuracy. The K-means clustering algorithm groups spectra based on the residuals between predicted and true structure descriptors, offering insights into which types of spectra tend to yield more accurate predictions. Similarly, PCA provides a way to visualize the spectra in terms of their principal components, with a scatter plot colored according to the performance of the corresponding structure descriptors. These unsupervised learning techniques offer an additional layer of analysis, enabling users to identify patterns in the data and understand how different features of the spectra influence the model's predictive power.

In summary, Block 4 offers a robust framework for both predicting structure descriptors and analyzing model performance. Through comprehensive statistical and graphical evaluations, along with the incorporation of unsupervised learning tools, it provides users with a thorough understanding of model reliability and predictive behavior, enhancing the overall interpretability and utility of the machine learning models within XASDAML.

## Toolkits (Module 4, 5 ,6)

XASDAML provides a suite of auxiliary toolkits for data visualization and statistical analysis, which, while not essential to the core workflow, significantly enhance users' understanding of the datasets. These modules are designed to facilitate the inspection, interpretation, and management of the data, thus supporting more informed decision-making in the analysis of XAS data.

### Toolkit 1：Data visualization (Module 4)

Module 4 provides a dedicated toolkit for visualizing spectra, wavelet data, and structure descriptors. This visualization is crucial for gaining an intuitive understanding of the data's key features, helping users to identify patterns, detect anomalies, and decide which datasets should be flagged as unphysical in the data filtering process. The module supports the visualization of a variety of datasets, including $\mu$ spectra, $\chi$ spectra, Fourier Transform (FT)/Reverse Fourier Transform (RFT) of $\chi$, wavelet data, and RDF for structural descriptors. Given the large volume of data, especially when working with wavelets, users can limit the range of datasets to be plotted by specifying start and end indices, as well as the interval (e.g., plotting every nth dataset). This option minimizes the number of plots and ensures that users can still capture the essential features of the data.

Additionally, Module 4 provides the ability to generate scatter plots comparing CN and CR from different methods within the same figure, allowing users to visually compare simulation performance across algorithms. Since the module runs within a Jupyter Notebook environment, users can easily modify and adapt the code to generate custom plots tailored to their needs. All outputs are automatically saved in timestamped folders, ensuring that results are organized and easily retrievable.

## Toolkit 2: Statistics for structure descriptors (Module 5)

Module 5 focuses on the statistical analysis of one-dimensional structure descriptor data. It provides a comprehensive set of statistical tools to help users gain a clear understanding of their data. These tools include essential metrics such as count, maximum, minimum, mean, median, and variance, all presented in a well-organized table for quick reference. This allows users to rapidly grasp the main characteristics of the dataset. The module also supports various visualization techniques to represent the data distribution. Histograms provide insight into the frequency distribution, helping users assess central tendencies, spread, and potential outliers. Density distribution graphs offer a smoothed representation of the data, allowing for the identification of underlying trends. Box plots summarize key statistics such as the median, quartiles, and outliers, offering a compact yet powerful tool for comparing the variability of different datasets.

In addition to these statistical summaries, Module 5 allows users to visualize the data before and after outlier filtering (Module 7) and dataset partitioning (Module 8), which can be particularly helpful for comparing the effects of these processes on the data. For datasets with continuous values, it is advisable to set a count step in the 'parameter set zone' to avoid arbitrary grouping of the data (default is 10), ensuring more meaningful visualizations. As with other modules, all outputs are automatically saved in timestamped folders for efficient result management. This module provides a comprehensive and flexible approach to understanding the statistical characteristics of the data, without altering the original datasets.

## Toolkit 3: Unsupervised ML analysis for spectra (Module 6)

Module 6 in XASDAML incorporates unsupervised ML techniques to assist in analyzing X-ray absorption spectra. The core functionality of this module is the visualization of data distributions, particularly for two-dimensional datasets. It integrates two widely-used unsupervised ML algorithms: K-means clustering and PCA. These algorithms are powerful tools for uncovering patterns in the data, reducing dimensionality, and identifying underlying structure.

K-means clustering is a simple yet efficient algorithm for grouping spectral data into clusters based on their similarity. It is particularly useful for large datasets and is widely applicable to various types of data. In Module 6, the elbow method is employed to determine the optimal number of clusters. This method involves calculating the intra-cluster sum of squared errors (SSE) for different numbers of clusters and plotting the results to identify a point (the "elbow") where adding more clusters does not significantly improve the model. This helps users visually select the ideal cluster count[28]. Once the optimal number of clusters is identified, the toolkit saves the cluster labels of each sample into a new CSV file and visualizes the spectral data by grouping them according to their respective clusters. The spectra of each group are displayed in separate graphs, allowing for easier comparison and analysis of clustering results. This approach helps researchers identify outliers and understand how different groups of data may represent distinct physical or chemical states of the material[29].

PCA is a technique used for dimensionality reduction, enabling the projection of complex, high-dimensional data into a lower-dimensional space. By retaining as much of the original variance as possible, PCA helps simplify data interpretation while preserving essential information. Module

6 visualizes the results of PCA by plotting the first two principal components, which generally explain more than 80% of the variance in most datasets. The Explained Variance Ratio is provided to indicate the proportion of variance captured by each principal component, giving users insight into how well the dimensionality reduction preserves the original data's information[29][30](as introduced in the support information). The results of PCA are displayed through scatter plots, where each point corresponds to a sample, and color coding can be applied based on various parameters, such as the clustering group or structural descriptors. This allows users to visually assess how well the spectra and structural descriptors align across the reduced dimensions. When combining PCA with K-means clustering, users can observe group structures and identify trends or anomalies that may not be apparent in the higher-dimensional space. The toolkit not only analyzes spectral distributions but also the relationship between the spectra and corresponding structural descriptors (such as CN and CR). After clustering the spectra with K-means, the toolkit uses box plots to display the distribution of structural data for each cluster, making it easy to compare the structural characteristics within and between clusters. This vertical comparison helps researchers explore how different groups of spectra are associated with different structural features. Moreover, after performing PCA, Module 6 enables horizontal comparisons between spectra and their corresponding structural descriptors. By plotting the first two principal components in a scatter plot, users can observe how different spectral patterns (e.g., $\mu$ and $\chi$ spectra) relate to variations in structure. The color coding of the scatter plot can reflect the clustering results or structural data, providing insights into how well the first two principal components explain the relationship between spectra and structure. For example, when using K-means clustering on the scatter plot, researchers can visually inspect the data for significant groupings or abnormal data points, aiding in the interpretation of complex spectral data and the identification of outliers.

  K-means clustering and PCA are widely used a lot in data processing and analysis. For example, PCA is commonly employed in data reduce dimension, extract key features in supervised ML, simplify the complexity of spectral data, and aid scientists in distinguishing between different samples and identifying information of structure descriptor. Besides, clustering the absorption spectrum data using the k-means method can help researchers identify the difference in groups of the data, making it easier to identify outliers and understand the chemical state of the material[29];
It is used for data preprocessing and identify material physical properties of XAS spectra when combined with relevant physical information provides feature references for subsequent machine learning[30].

  The visualizations provided by PCA and K-means clustering in Module 6 also serve as a valuable resource for feature selection when building machine learning models. By observing how the principal components or clusters relate to structural descriptors, users can gain a deeper understanding of which features are most relevant for training the ML models[31]. This can help guide the selection of appropriate features, enhancing the performance and interpretability of predictive models built on XAS data. Similar to Module 5, Module 6 allows users to analyze datasets at different stages of preprocessing, such as before and after outlier screening (via Module 7) and data partitioning (via Module 8). This flexibility enables users to compare how the spectral and structural distributions evolve after these steps, facilitating the refinement of data before building predictive models. The toolkit can process both CSV datasets (prior to partitioning) and TXT datasets (after partitioning) to ensure that users can perform analysis on datasets at various stages, including the training, validation, and test sets.

In summary, Module 6 in XASDAML offers a comprehensive set of tools for unsupervised analysis of X-ray absorption spectra. By integrating K-means clustering and PCA, this module allows users to explore the distribution of spectra, investigate relationships between spectra and structural descriptors, and gain valuable insights for model building. The visualization of clustering and principal components aids in understanding complex data structures, identifying outliers, and selecting relevant features for subsequent machine learning tasks. These capabilities enhance the overall workflow, helping researchers better interpret their data and improve the performance of ML models.

## Application

Copper is a common metal that popular used in electrochemistry material, its foil is in the indispensableo standards at the XAS baemlines that used to to calibrate the absorption edge of XAS in the measurement. We take it here as a candidate to see what we can get from the XAS spectra of copper with XASDAM, and it is also a good example to show the performance of this ML-based XAS data process framework.

We started from the construction of XAS datasets for copper. 5000 different geometric configurations of copper is derived from molecular dynamics (MD) ,where each contains 108 copper atoms, more details can be found in SI section X. As we consider here only the coordinate number and average bond length of fisrt shell around absorber, 5000 datsets are sufficient to get a good analysis accuracy, as we will see later. Note that more datasets may bevneeded in the more detailed geometric structure analysis, such as the bond angle of coordinates.

With 5000 atomic structure of copper system, we put them into module 1 of XASDAML and got their XAS spectra, we release most the parameters in the module unchanged (default values) except 'num_processes' was set to be 10 as XASDAML runs at a server with an Intel(R) Xeon(R) Gold 6230R CPU @ 2.10GHz which has a thread of 96, the simulation finished in 17 hours and 35 mins, by comparion, more than 164 hours are needed for unparallel XAS simulation for 5000 structure with FEFF as ordinary cost for a 100 atoms with this package is about 4 minsiutes.

We put 5000 geometric structures of copper system into module 2 and got their descriptor of structure, the coordinate number CN, the average bond length CR, of fisrst shell around absorber, and radial distrubtion function RDF, as shown in Table SI1 and Table SI2 in the supporting infroamtion section'Cluster analusis and PCA'. We don't know which alforithms are better to identify the neighbouring atoms, so all of them are ued here in the simulation ('methods= 'EconNN, VoronoiNN', ' JmolNN' ,'MinimumDistanceNN', ' CrystalNN'. ' BrunnerNN_relative').

Once the spectrum-structure datasets were derived, we were anxious to see what spectra we have obtained and wonder whether they are reasonable or not in profiles, so we took module 4 to read the simulated spectra output by module 1 and output their figures. As the printout of one plot for all the spectra($\mu$) to copper system with one configuration takes about 9 minutes and 51 seconds, we set 'start' and 'interval' to be '500' and '100', so only 46 plots($\mu$) were printout to save time,which takes about 5.49 seconds. Some of them are shown in figure 3(c). Note that the system configurations of copper derived from MD are very similar, it is not necessary to printout all the plots of spectra for all the configuations. There always a trade-off between the expectation of seeing more spctra profiles and less time for excitation of module 4.

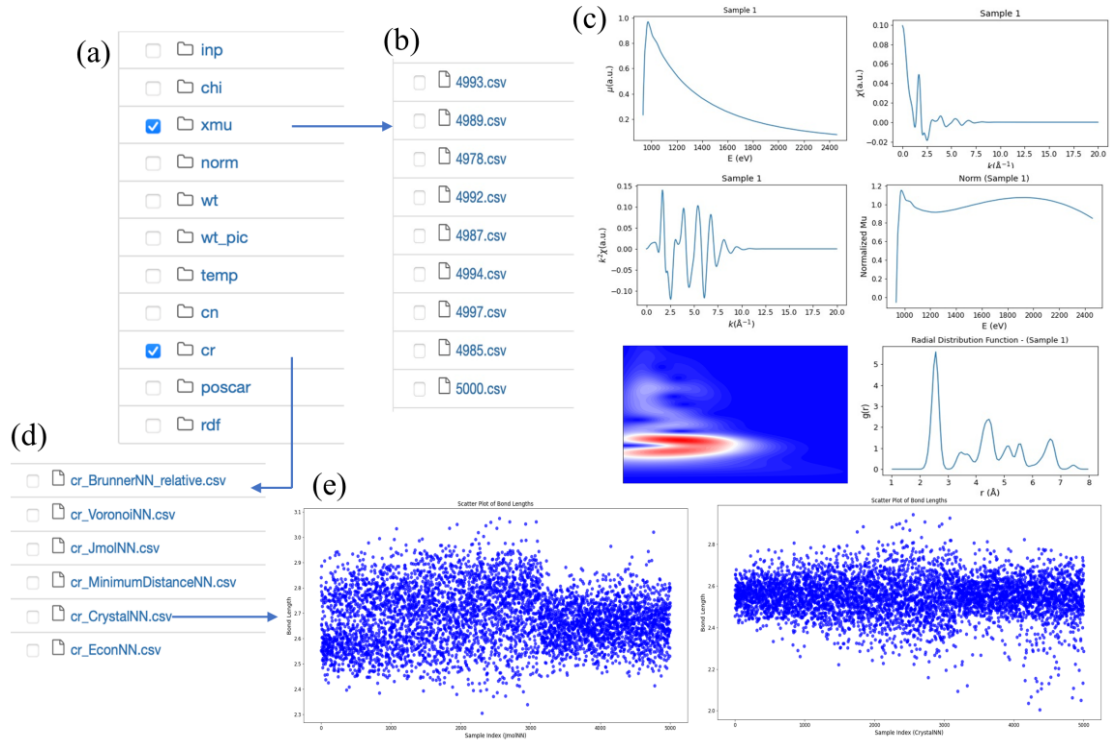

Figure 3. Dataset construction and visualization of the Cu system in XASDAML. (a) Calculation of absorption spectra and feature extraction, demonstrating the generation of $\mu$ spectra and the extraction of key spectral features. (b) Data storage display, illustrating the organized storage and structure of the generated datasets within the XASDAML framework. (c) Visualization of extracted absorption spectrum features($\mu$, $\chi$, $k^2\chi$, norm, wt, RDF). (d) Results from different bound length calculation methods. (e) Statistical distribution of calculated CR using the JmolNN and CrystalNN methods, illustrating the distribution patterns of structural parameters within the Cu system.

After obtaining the spectral and structural data, the Statistical and Data Analysis Module was employed to conduct a detailed statistical analysis of the Cu dataset, ensuring a comprehensive understanding of the distribution characteristics of CN and CR. We performs statistical computations on the raw dataset, including sample size, mean, median, quartiles, variance, maximum, and minimum values. Additionally, it automatically generates corresponding visualizations, such as bar plots, density plots, and box plots. By analyzing the statistical distributions of CN and CR, potential structural anomalies can also be identified. Figure 4 presents the distributions of CN and CR obtained using different neighbor identification algorithms. These visualizations and tabular data allow for the rapid detection of outliers or distribution deviations within the dataset, which in turn informs the selection of appropriate filtering thresholds and data selection criteria for subsequent modeling. As a result, the prediction models trained on this dataset can achieve higher accuracy and stability.

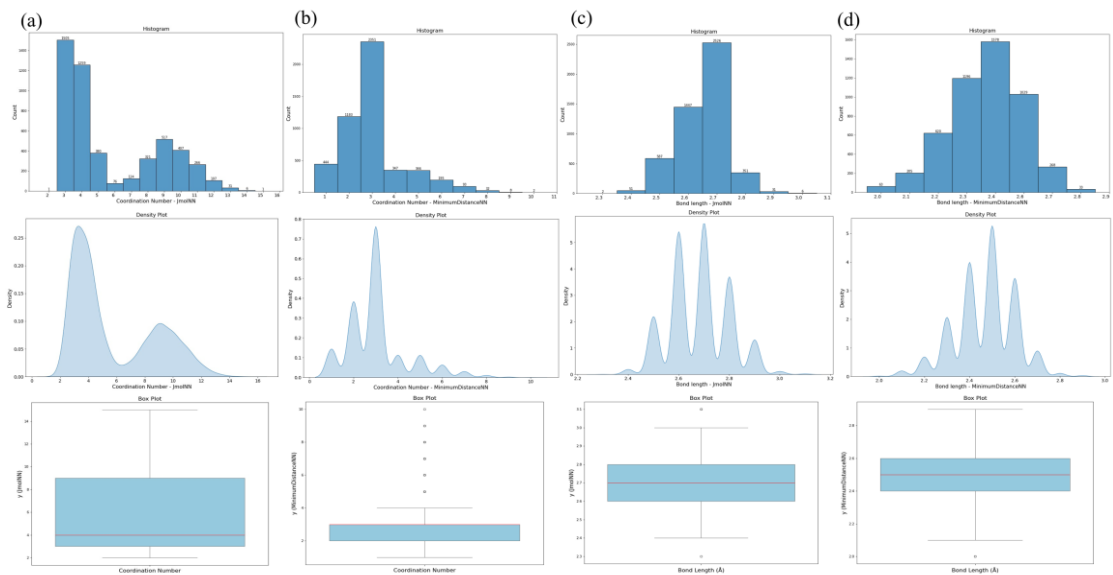

Figure 4. Histograms, box plots, and density distributions of CN and CR in the Cu system, calculated using different nearest neighbor methods. (a) CN using JmolNN; (b) CN using MinimumDistanceNN; (c) CR using JmolNN; and (d) CR using MinimumDistanceNN.

To further explore the correlation between absorption spectra and structural descriptors, as well as uncover intrinsic patterns within the spectral dataset, the Data Analysis of Spectra and Structure Descriptors Module was utilized, applying PCA and clustering algorithms. Taking $\mu$ spectra as an example, PCA was first performed to reduce dimensionality and visualize the distribution of the first two principal components in a scatter plot. Typically, the cumulative variance contribution of the first two components exceeds 90% (as shown in Figure 5), indicating that most spectral variations can be captured by just a few key principal components. Following this, the K-means clustering algorithm was applied to categorize the $\mu$ spectra. Initially, the dataset was divided into three clusters, and their corresponding spectra were overlaid in a single plot (Figure 5(a, b)) to facilitate the rapid identification of potential outlier spectral types. The dataset was then further classified into four clusters using the Elbow Method, and the $\mu$ spectra of each category were analyzed in aggregate. Additionally, box plots were used to display the distribution of CN and CR for each cluster (Figure 5(c)). By comparing these box plots with previous statistical distribution visualizations, it was observed that using the JmolNN algorithm for neighbor identification better suited the system's structural characteristics. Furthermore, $\mu$ spectra demonstrated significantly higher sensitivity to CN than to CR.

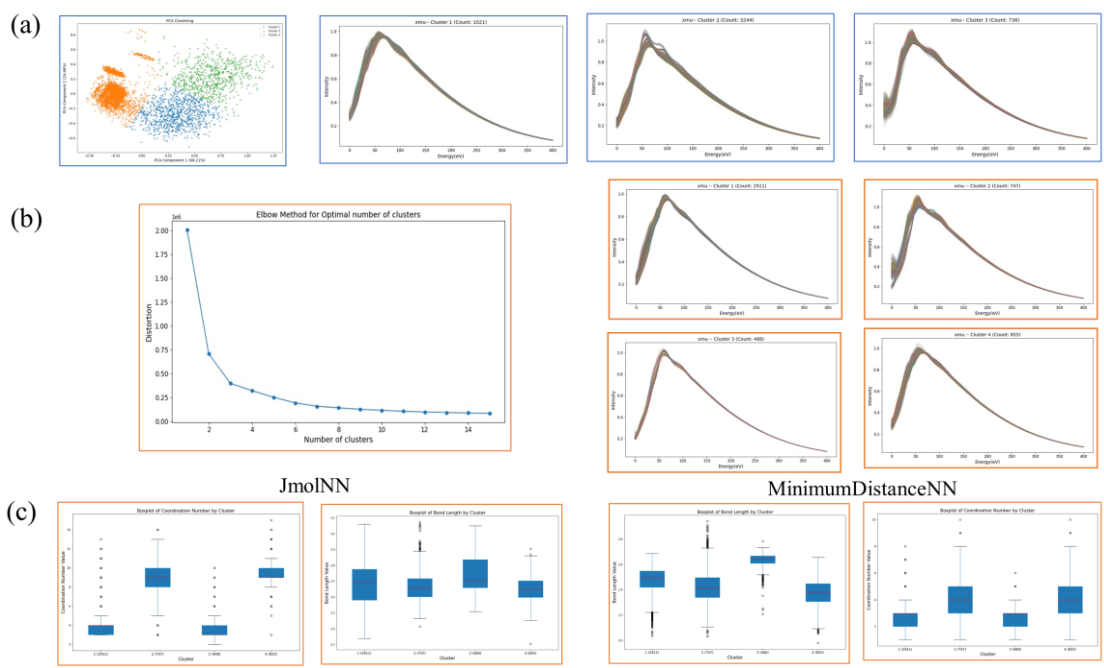

Figure5. Clustering analysis of the Cu system before outlier removal. (a) Scatter plot of the first two principal components obtained from PCA applied to $\mu$ spectra, with data points grouped into three clusters using the K-means algorithm. (b) Identification of four K-means clusters for $\mu$ spectra using the elbow method. (c) Box plots illustrating the distribution of CN and CR, calculated using JmolNN and MinimumDistanceNN methods, within the four K-means clusters of $\mu$ spectra.

Following the preliminary analysis of the relationships between spectral features and structural descriptors, the Filtering Module was applied to remove unreasonable data points, thereby improving data quality. The filtering criteria were set such that CN values were restricted to the range of 2–12 and CR values to 2.4–3.0 Å. Any samples falling outside these thresholds were excluded from further analysis. Using structural descriptors computed via the Jmol algorithm, 38 samples were removed due to exceeding the CN limit, while 26 samples were filtered out for CR deviations, resulting in a total of 64 samples being discarded from the original 5000 dataset (Figure 6(a)). After filtering, the remaining spectral and structural data were partitioned into training, validation, and test sets in a 7:2:1 ratio. The processed datasets were subsequently saved in TXT format for future modeling and analysis (Figure 6(b)).

To further investigate the relationship between spectral features and CN or CR, PCA analysis was applied to the filtered dataset. Similar to the previous steps, PCA was performed on the $\mu$ spectra, and the principal component scores were visualized in a scatter plot. In this case, the complete dataset was used (by specifying all available samples in the parameters) to provide a more comprehensive representation of the results. To evaluate the correlation between the principal components and CN or CR, the scatter plot was color-coded according to different CN or CR values (Figure 6(c)), allowing for the observation of how the first two principal components influence subsequent prediction accuracy. From Figure 6(c), it can also be observed that $\mu$ spectra exhibit significantly higher sensitivity to CN than to CR. When coloring the data based on CN values, the points are more distinctly separated compared to when using CR, indicating that CN has a stronger influence on spectral variations. This further supports the earlier findings that $\mu$ spectra are more effective in capturing CN-related structural differences. After completing PCA, additional preprocessing steps such as standardization or PCA transformation could be applied if necessary (Figure 6(d)). These preprocessing operations and their parameters were recorded to ensure consistency and traceability in subsequent model training.

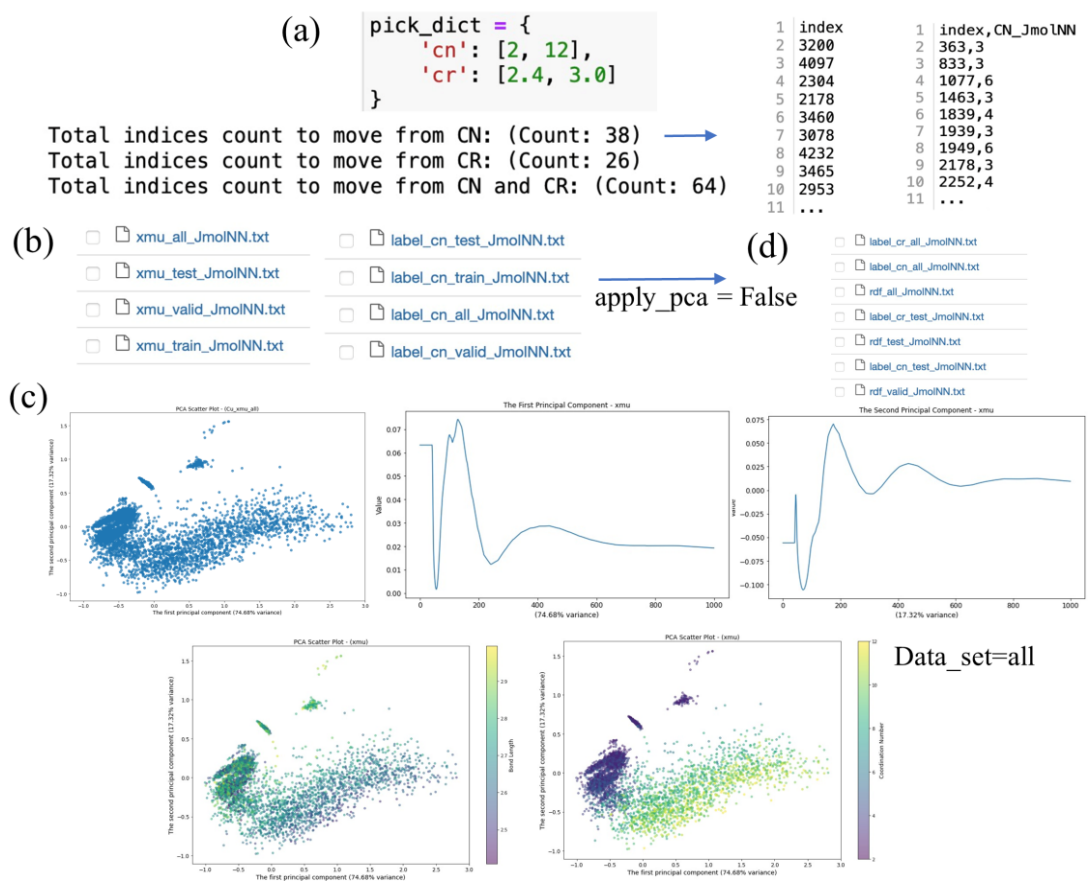

Figure 6. Data analysis and dataset standardization after partitioning the Cu system dataset. (a) Outlier filtering applied to the original Cu dataset, removing 64 samples outside the predefined ranges of CN (2–12) and CR (2.4–3.0) based on the Jmol algorithm. (b) Splitting dateset into training, validation, and test sets in a 7:2:1 ratio, with the partitioned datasets saved as TXT files. Both original and interpolated spectra are plotted for comparison. (c) PCA of the $\mu$ spectra, displaying the first two principal components in a scatter plot. Samples are color-coded according to their CN and CR to investigate the influence of spectral characteristics on prediction results. (d) Application of the standardization module, demonstrating data standardization prior to neural network modeling.

In the preceding steps, statistical analysis, clustering, and PCA were used to explore the distribution characteristics of the dataset and examine the correlation between absorption spectra and structural descriptors. Following this exploratory analysis, a machine learning model was trained to enable rapid prediction of structural parameters from spectral data. After dataset partitioning and standardization, the Machine Learning Module provided three commonly used supervised learning models: MLP, CNN, and RF. For this study, the classic MLP model was selected. The training input consisted of $\mu$ spectra, while the corresponding CN and CR were used as the model's output. During training, the system automatically generated and saved the model architecture, training and validation curves, and the optimized model parameters, ensuring traceability for future model comparisons or reapplications. In the prediction phase, the Prediction Module allowed the pre-trained model parameters to be directly loaded onto new data for inference. According to the experimental results, the CN prediction achieved an $R^2$ of 0.95 with a MSE of approximately 0.37 (Figure 7). In contrast, the CR prediction demonstrated relatively lower accuracy, with an $R^2$ of approximately 0.58 and an MSE of 0.05 (Figure SI 1). As shown in Figure SI 2, a dataset size of 5000 was sufficient to meet the model's convergence requirements. To facilitate a clearer comparison between predicted and actual values, line charts were generated for CN and CR (Figures 7 and SI 1), enabling visual assessment of prediction accuracy based on $R^2$ values and plotted curves. The results indicate that $\mu$ spectra exhibit greater sensitivity to CN than to CR, corroborating previous spectral-structural correlation analyses. If necessary, the module also provides options for rounding predicted values or applying additional post-processing; however, to maintain a direct comparison of CN and CR prediction accuracy, no rounding was applied in this case.

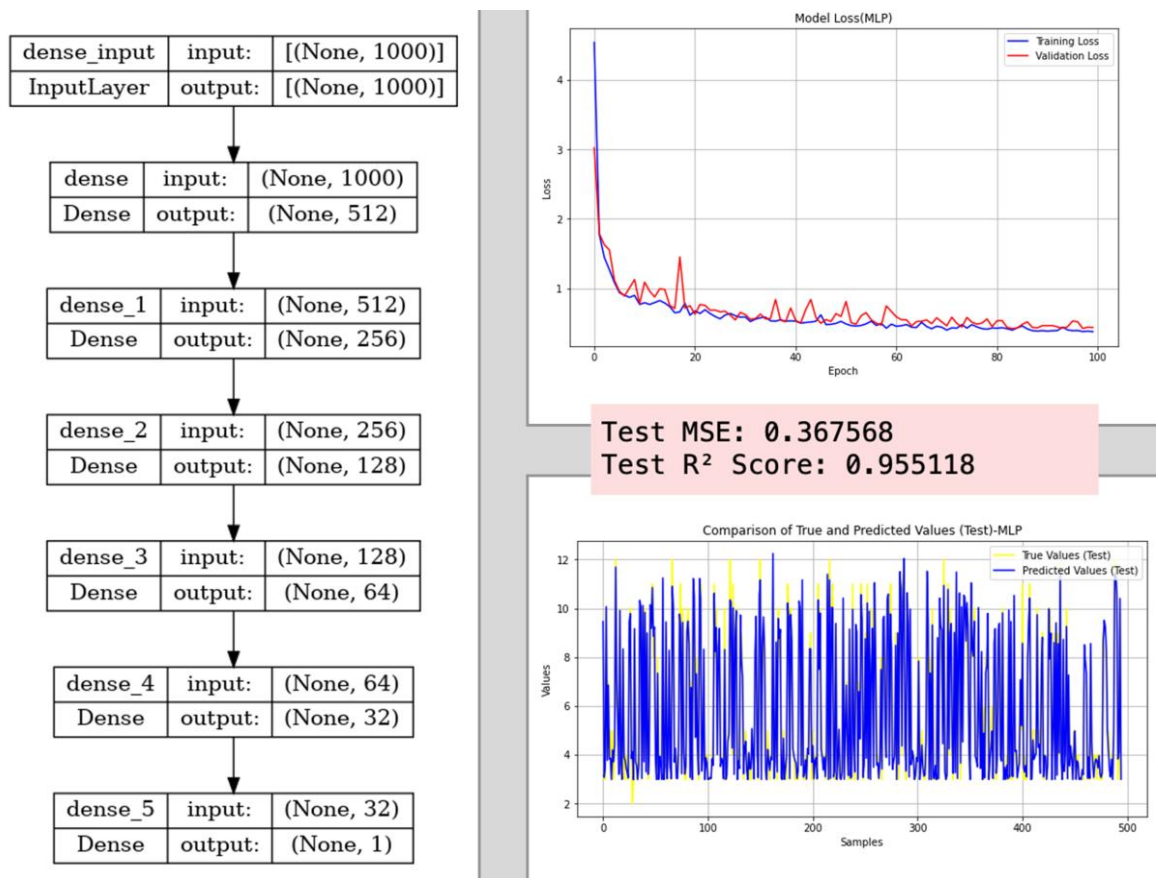

Figure 7. Prediction of CN in the Cu system using the MLP model, illustrates the prediction accuracy and the corresponding error metrics for the model's performance on the Cu dataset.

After obtaining the machine learning prediction model and its corresponding results, the Prediction Result Analysis Module can be used to evaluate model performance and analyze error distribution. By directly comparing the predicted and actual values of CN, along with the corresponding $\mu$ spectra, the deviation between the model's predictions and the true structural properties can be quickly assessed. Figure 8 presents key evaluation metrics, including mean, variance, MAE, MSE, and $R^2$, providing a reliable basis for quantifying prediction accuracy and stability. Beyond these fundamental statistical indicators, additional visualizations and interpretations of prediction errors can be achieved using the Statistical and Data Analysis Module. For example, applying K-means clustering to prediction errors or performing PCA on the error distribution allows for the identification of spectral types that are more prone to large deviations, helping pinpoint potential anomalies or areas for model refinement. Figure 9(a) compares the predicted and true CN values using a bar chart, supplemented by scatter and line plots to provide a more intuitive representation of the model's predictive accuracy. Figure 10(a, c) illustrates the error distribution across different spectral clusters obtained via K-means clustering, along with their corresponding $\mu$ spectra, aiding in the identification of spectral features most strongly associated with large prediction errors. Similarly, Figure 9(b) and Figure 10(b) combine bar and box plots, with the PCA coordinate system color-coded according to prediction errors, facilitating a detailed examination of error patterns or anomalies within specific clusters or principal components.

By integrating clustering and PCA analysis into the evaluation of machine learning model predictions, a more precise assessment of model performance can be achieved while uncovering deeper relationships between structural descriptors and spectral features. This comprehensive visualization and analytical approach is instrumental in guiding subsequent model refinements, such as re-evaluating feature selection strategies or adjusting threshold filtering criteria. Examining the relationships between prediction errors, spectral data, and structural parameters provides a more profound understanding of the complex mapping between spectra and structures in the Cu system, offering valuable insights for future spectral analysis in other material systems.

```
Analysis Report for Predictions and True Values
================================================
1. Mean:
    - Mean of predictions: 5.758483033932135
    - Mean of true values: 5.746506986027944

2. Variance:
    - Variance of predictions: 7.756040812586403
    - Variance of true values: 8.033545683084927

3. Outliers:
    - Outliers in predictions: [12.0, 12.0, 12.0, 12.0]
    - Outliers in true values: [13.0, 13.0, 13.0, 13.0, 12.0, 12.0, 12.0, 13.0]

4. Error Metrics:
    - Mean Absolute Error (MAE): 0.3473053892215569
    - Mean Squared Error (MSE): 0.4111776447105788
    - Root Mean Squared Error (RMSE): 0.6412313503803279
    - Mean Absolute Percentage Error (MAPE): 0.05690762442259448
    - Explained Variance: 0.9488352671159099
    - Max Error: 2.0
    - R² Score: 0.9488174137633478
```

Figure 8. Prediction result analysis of the Cu system, including an analysis report comparing predicted and true values with metrics such as mean, variance, outliers, and error metrics.

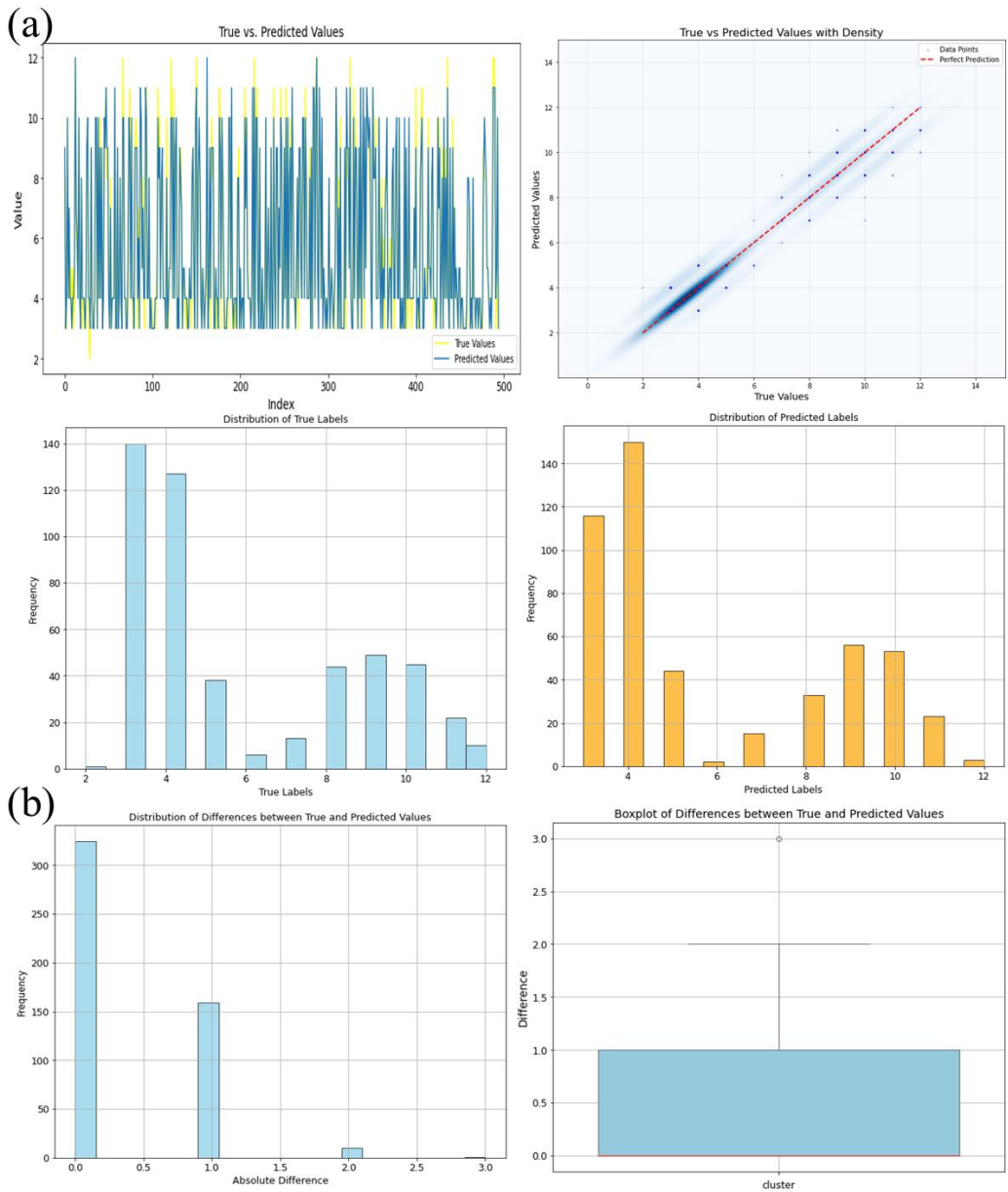

Figure 9. Prediction Result for Cu System Using XASDAML. (a) Bar chart comparing predicted and true CN values, alongside parity plots and line charts, illustrating the model's predictive accuracy. (b)Bar chart and box plot of prediction errors.

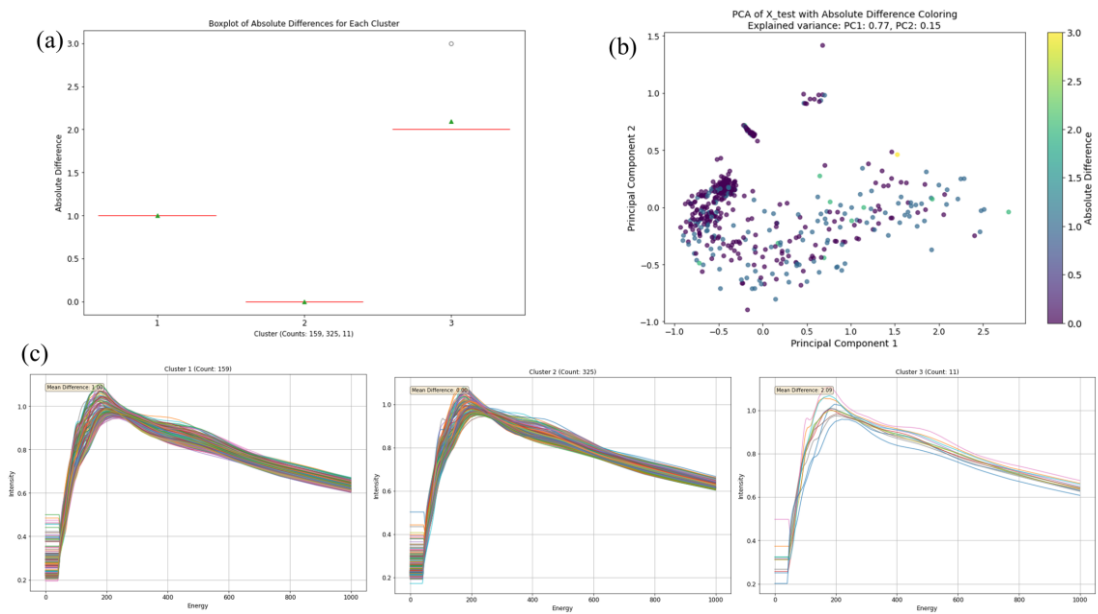

Figure 10. Prediction result analysis for the Cu system using XASDAML. (a) K-means clustering of prediction errors across different data points. (b) PCA analysis of $\mu$ spectra, with samples colored according to their prediction errors, providing an intuitive assessment of the model's overall performance. (c) $\mu$ spectra plots for each cluster obtained from the K-means clustering of prediction errors.

## Discussion

XASDAML is built with a modular architecture that provides flexibility and adaptability in processing and analyzing XAS data. A key strength of this framework is its ability to replace or upgrade individual modules, as long as the input and output files conform to a consistent format. This design enabling users to customize the system to their research needs or incorporate future advancements in XAS data analysis without disrupting the overall workflow.

Many XAS analyses benefit from enhanced feature extraction[32], where researchers derive more detailed descriptors from the raw absorption spectra to strengthen subsequent machine learning steps. Although XASDAML currently uses wavelet transforms as its primary method of spectral feature extraction, we have reserved a placeholder in Block 1 for a more comprehensive feature descriptor module. This planned module could incorporate additional techniques—such as polynomial fitting, the cumulative distribution function, or higher-order transformations—to generate richer inputs for machine learning models. These enhancements would offer richer inputs for machine learning models, deepening the analysis of spectral-structural relationships.

Graph Neural Networks (GNNs) have emerged as a powerful solution for modeling data that exhibit intricate relational structures, making them well-suited to cases where local and interatomic interactions play critical roles. XAS data inherently involves complex interatomic relationships and local structural information, which are well-suited for modeling with graph-based approaches. Preliminary applications of GNNs to XAS-related problems have shown a significant advantage in capturing intricate structural correlations[33]. Notably, we have already applied GNNs to XAS-related work, and our preliminary results demonstrate their significant advantage in capturing complex structural correlations[34]. Integrating GNNs into XASDAML would not only improve predictive performance but also broaden the framework's applicability for materials design and discovery.

While XASDAML has shown promising results with simulated data, applying it to experimental datasets involves bridging the significant quantitative mismatch between theoretical calculations and real measurements—beyond capturing overall structural features, the accuracy of simulated spectra depends heavily on the precision of the computational software. Meanwhile, secondary factors, such as instrument calibration errors and varying signal-to-noise ratios, can further complicate model training. To address these challenges, several strategies can be employed. This helps train the model to recognize patterns even in noisy or distorted data, improving its robustness and generalization to experimental datasets. Preprocessing steps like normalization, standardization, or spectral energy shift may be necessary to align feature, given that experimental spectra can vary in baseline intensity or absorption edge shifts. Techniques such as z-score normalization or baseline correction can help standardize these differences, boosting model accuracy. Finally, ensuring that the model can handle the variability in experimental conditions—such as differences in synchrotron facilities—is crucial, and strategies like cross-validation across multiple setups or incorporating diverse training datasets help achieve robust generalization. Crucially, if the theoretical spectra are reasonably accurate in reflecting key structural relationships, the machine learning approach offered by XASDAML can still facilitate large-scale, automated parsing of spectral–structure correlations and thereby extend its utility to experimental data.

## Conclusion and Outlook

This study introduced XASDAML, a novel data processing framework that applies machine learning algorithms to streamline the analysis and prediction of XAS data. The framework integrates multiple stages, including spectral and structural calculations, data visualization, statistical analysis, model training, prediction, and result interpretation. By combining machine learning methods with traditional XAS data processing, XASDAML enhances data handling efficiency and offers deeper insights into the correlation between spectral and structural properties. The modular design and open-source nature of the framework ensure broad applicability to a variety of datasets and research objectives, while the Jupyter Notebook interface facilitates ease of use and encourages reproducibility.

The application of XASDAML to a Cu dataset demonstrated its effective handling of large and complex data sets. By employing statistical methods such as PCA and clustering, the framework successfully identified significant patterns, while the machine learning models accurately predicted CN and CR based on spectral features like $\mu$ spectra. These outcomes highlight the platform's ability to guide researchers in selecting appropriate structural calculation methods and extracting meaningful structure-property relationships.

Compared to previous XAS-ML tools, XASDAML offers enhanced operational flexibility and customization by unifying all stages of data handling, thereby eliminating the need for multiple disparate tools. This streamlined approach lowers the barrier to entry for users with varying levels of expertise in integrating machine learning methods into the XAS data processing workflow and supports extensive customization and scalability to meet diverse research requirements.

Additionally, XASDAML provides comprehensive features for data exploration, visualization, and model selection, enabling researchers to derive more precise physical and chemical insights. Its open-source nature fosters collaboration and continuous improvement, setting it apart from earlier frameworks by offering a more integrated and adaptable solution for advanced XAS data analysis.

Despite these advancements, challenges remain in XASDAML. Key areas for improvement include extracting more accurate and meaningful physical information from noisy or complex data and enhancing the framework's generalization capabilities across different structural datasets. Moreover, the robustness of machine learning models and their integration with physics knowledge—including physical constraints, enhanced preprocessing, and uncertainty quantification—will be essential to ensure reliable and precise predictions.

Future development will focus on incorporating more advanced machine learning algorithms, refining model interpretability, and improving data preprocessing and denoising techniques. With these enhancements, XASDAML has the potential to become an even more powerful tool for materials characterization, enabling more precise and efficient XAS data analysis.

## CRediT authorship contribution statement

**Xue Han**: Writing – original draft, Visualization, Validation, Methodology, Investigation, Formal analysis, Data curation, Conceptualization. **Haodong Yao**: Writing – original draft, Writing – review & editing, Validation, Methodology, Investigation, Formal analysis, Conceptualization. **Fei Zhan**: Software, Data curation, Investigation. **Xueqi Song**: Software. **Junfang Zhao**: Supervision, Project administration, Writing – review & editing. **Haifeng Zhao**: Conceptualization, Supervision, Project administration, Funding acquisition, Writing – review & editing.

## Declaration of Competing Interest

The authors declare that they have no known competing financial interests or personal relationships that could have appeared to influence the work reported in this paper.

## Acknowledgements

We acknowledge financial support from the National Key Program of China (2020YFA0405800) and Platform of Advanced Photon Source Technology (PAPS), one of the Key Projects in the Planning of Huairou National Comprehensive Science Center, Beijing.

## Code available

The data and code supporting this work are available at the GitHub repository (https://github.com/BSRF-XA/XASDAML).